\documentclass[a4paper,UKenglish,cleveref, autoref, thm-restate]{lipics-v2021}

\usepackage[table]{xcolor}
\usepackage{tcolorbox}
\usepackage[frozencache]{minted}
\usepackage{minted}
\def\us{\char`\_}
\def\coqin#1{\mintinline{ssr}{#1}}
\usepackage{syntax}
\usepackage{tabularx}
\usepackage{stmaryrd}
\usepackage{simplebnf}
\usepackage{comment}
\usepackage{rotating}
\usepackage{tablefootnote}
\usepackage{booktabs}
\usepackage{cite}

\newcommand{\Real}{{\mathbb R}}
\newcommand{\Pp}{{\mathcal P}}
\newcommand{\Sp}{{\cal S}}

\newcommand{\losssymbol}{{\cal L}}
\newcommand{\lossfn}{\losssymbol}

\newcommand{\naryConj}{\bigwedge\nolimits_M}
\newcommand{\naryDisj}{\bigvee\nolimits_M}

\def\DL{\textrm{DL}}
\def\DLtwo{\textrm{DL2}}
\def\STL{\textrm{STL}}
\def\product{\textrm{product}}
\def\Yager{\textrm{Yager}}
\def\Luka{\textrm{Łukasiewicz}}
\def\Godel{\textrm{Gödel}}
\def\True{\textsl{True}}
\def\False{\textsl{False}}

\newcommand{\tDLtwo}[1]{\llbracket #1 \rrbracket_{\DLtwo}}
\newcommand{\tGodel}[1]{\llbracket #1 \rrbracket_{\textrm{G}}}
\newcommand{\tSTL}[1]{\llbracket #1 \rrbracket_{\STL}}
\newcommand{\tlukasiewicz}[1]{\llbracket #1 \rrbracket_{\textrm{Ł}}}
\newcommand{\tyager}[1]{\llbracket #1 \rrbracket_{\textrm{Y}}}
\newcommand{\tproduct}[1]{\llbracket #1 \rrbracket_{\textrm{P}}}
\newcommand{\tdl}[1]{\llbracket #1 \rrbracket_{\DL}}
\newcommand{\tbool}[1]{\llbracket #1 \rrbracket_{\textrm{B}}}
\newcommand{\tempty}[1]{\llbracket #1 \rrbracket}

\newcommand{\tdlbig}[1]{\bigg\llbracket #1 \bigg\rrbracket_{\DL}}

\newcommand{\lam}[1]{\lambda #1\ .}

\newcommand{\fold}{\text{fold}}

\newcommand{\vect}[1]{\mathbf{#1}}
\newcommand{\x}{\vect{x}} 
\newcommand{\ve}{\vect{v}} 
\newcommand{\y}{\vect{y}} 

\newcommand{\Nat}{\mathbb{N}}

\newcommand{\pval}{p}

\newcommand{\VecType}[1]{\ensuremath{\text{Vec } #1}}
\newcommand{\FinType}[1]{\ensuremath{\text{Index } #1}}
\newcommand{\BoolType}{\ensuremath{\text{Bool}}}
\newcommand{\RealType}{\ensuremath{\text{Real}}}
\newcommand{\FunType}[2]{\ensuremath{\text{Fun } #1 \text{ } #2}}

\newcommand*{\impl}{\Rightarrow}

\title{Taming Differentiable Logics with Coq Formalisation} 

\author{Reynald Affeldt}{National Institute of Advanced Industrial Science and 
Technology (AIST)
}{reynald.affeldt@aist.go.jp}{https://orcid.org/0000-0002-2327-953X}{}

\author{Alessandro Bruni}{IT-University of Copenhagen, Denmark}{brun@itu.dk}{https://orcid.org/0000-0003-2946-9462}{}

\author{Ekaterina Komendantskaya}{Southampton University, UK \and 
Heriot-Watt University, 
UK}{e.komendantskaya@soton.ac.uk}{https://orcid.org/0000-0002-3240-0987}{}

\author{Natalia Ślusarz}{Heriot-Watt University, UK} 
{nds1@hw.ac.uk}{https://orcid.org/0000-0001-5729-9208}{}

\author{Kathrin Stark}{Heriot-Watt University, 
UK}{k.stark@hw.ac.uk}{https://orcid.org/0000-0002-7086-6518}{}

\authorrunning{Affeldt, R. et al.} 

\Copyright{Reynald Affeldt, Alessandro Bruni, Ekaterina 
Komendantskaya, Natalia Ślusarz, Kathrin Stark} 



\category{} 

\relatedversion{} 
\supplementdetails[subcategory={Source 
Code}]{Software}{https://github.com/ndslusarz/formal_LDL}

\nolinenumbers 

\EventEditors{Yves Bertot, Temur Kutsia, and Michael Norrish}
\EventNoEds{3}
\EventLongTitle{15th International Conference on Interactive Theorem Proving 
(ITP 2024)}
\EventShortTitle{ITP 2024}
\EventAcronym{ITP}
\EventYear{2024}
\EventDate{September 9-14, 2024}
\EventLocation{Tbilisi, Georgia}
\EventLogo{}
\SeriesVolume{309}
\ArticleNo{34}

\newtcolorbox{authorComment}[1]{colback=#1}

\usepackage[textwidth=2.5cm,textsize=footnotesize,colorinlistoftodos]{todonotes}

\def\coq{\textsc{Coq}}
\def\mathcomp{\textsc{MathComp}}
\def\analysis{\textsc{MathComp-Analysis}}

\begin{document}

\maketitle

\begin{abstract}
For performance and verification in machine learning, new
methods have recently been proposed that optimise learning
systems to satisfy formally expressed logical properties.
Among these methods, differentiable logics (DLs) are used to
translate propositional or first-order formulae into loss
functions deployed for optimisation in machine learning.
At the same time, recent attempts to give programming language support
for verification of neural networks showed that DLs can be used to
compile verification properties to machine-learning backends.  This
situation is calling for stronger guarantees about the soundness of
such compilers, the soundness and compositionality of DLs, and the
differentiability and performance of the resulting loss functions.  In
this paper, we propose an approach to formalise existing DLs using the
Mathematical Components library in the Coq proof assistant.  Thanks to
this formalisation, we are able to give uniform semantics to otherwise
disparate DLs, give formal proofs to existing informal arguments, find
errors in previous work, and provide formal proofs to missing
conjectured properties.  This work is meant as a stepping stone for
the development of programming language support for verification of
machine learning.
\end{abstract}

\def\newterm#1{{\sl #1}}
\def\mydef{\overset{\textrm{def}}{=}}

\section{Introduction}

This work aims to contribute to the field of formal verification of
artificial intelligence, more precisely machine learning, i.e., the
study of algorithms that learn statistically from data. Neural networks
are the most common technical device used in machine learning.
The standard learning algorithms (such as gradient descent) use a
\newterm{loss function}
$\losssymbol: \Real^m \times \Real^n \rightarrow \Real$ to optimise
the network's parameters (say, $\theta$) to fit the input-output
vectors given by the data in a way that the loss $\losssymbol(\x, \y)$
is minimised. This optimisation objective is usually denoted as
$ \min_{\theta} \lossfn(\x, \y) $. 

Most approaches to verification of neural networks consist of an
automated procedure based on SMT solving, abstract interpretation, or
branch-and-bound techniques (see, e.g., Albarghouthi's
survey~\cite{albarghouthi-book}).
Verification typically applies after training because traditional
learning is purely data-driven and thus agnostic to verification
properties.
In contrast, \newterm{property-guided training} takes place once the
verification properties are stated.
More precisely, verification of neural networks consists of two parts:
statement and verification of a given property, and training
of the neural network that optimises the neural network's parameters
towards satisfying the given property.

However, naively or manually performed mapping of a logical property
to an optimisation task results in major discrepancies (as shown by
Casadio et al.~\cite{CasadioKDKKAR22}). This suggests the need to have
tools for property-guided training. One approach is to provide
programming language support for property-driven development of neural
networks that involves specification, verification, and optimisation in
a safe-by-construction environment.
As an example, Vehicle~\cite{FoMLAS2023,daggitt2024vehicle} provides this 
support in the form of
a Haskell DSL, with a higher-order typed specification
language, in which required neural network properties can be clearly
documented, and type-driven compilation which can take care of
correct-by-construction translation of properties into both the
language of neural network solvers and loss functions.

To generate loss functions from a logical property, one can use
\newterm{Differentiable Logics} (DLs).
Well-studied \newterm{fuzzy logics} that date back to the works of
Łukasiewicz and G\"{o}del can be used as DLs~\cite{van2022analyzing}.
Recently, both verification and machine-learning communities
formulated alternative DLs such as \DLtwo~\cite{fischer2019dl2} and
\STL~\cite{varnai}; the latter was shown to 
be more performant in
optimisation tasks. These DLs are very different; for example, they do
not agree on the domains of the resulting loss functions: fuzzy logics
have domain $[0,1]$,
the domain of \DLtwo{} corresponds to the Lawvere quantale
$(- \infty, 0]$, and
\STL{}'s domain is $(-\infty, +\infty)$ (all intervals are equipped with
the usual ordering on reals).  Each domain has a designated value for
truth (e.g., $1$ in fuzzy logics, $0$ in \DLtwo{}, $+\infty$ in \STL)
and falsity ($0$, $-\infty$, $-\infty$, respectively).

Vehicle uses DLs to translate logical properties into loss functions. In order to 
ensure the correctness of the translation, a DL needs to satisfy a number of properties:
\begin{itemize}
\item \newterm{Soundness}: if a property interprets as ``true'' in the
  chosen DL domain, then it is true in the boolean logic, and
  similarly for false.
\item \newterm{Compositionality}: the translation function should 
  preserve the structural properties, e.g.,
  (the translation of) negation should compose with conjunction and
  disjunction, and (the translation of) conjunction and disjunction
  should satisfy the usual properties of idempotence, commutativity,
  and associativity.
\item \newterm{Shadow-lifting}: the resulting functions should have
  partial derivatives that can characterise the idea of gradual
  improvement in training~\cite{varnai}. For example, a translation of
  a conjunction should evaluate to a higher value if the value of one
  of its conjuncts increases.
\end{itemize}

Unfortunately, none of the existing DLs satisfies all of these
requirements~\cite{ldl,varnai}.
Therefore, future tools and compilers such as Vehicle may need to
provide support for incorporating a range of DLs for different
scenarios.  

This conclusion brings to the forefront the need for a generic
framework in which logical and geometric properties of different DLs
can be formalised and proven.
In this paper, we propose a unified formalisation of DLs to lay down
the ground for the development of a reliable neural network verification tool.
For that purpose, we will build on top of previous work that has already proposed
a common presentation of DLs~\cite{ldl}.
In order to handle the verification of translation from properties to
loss functions, we use the \coq{} proof assistant in which numerous
formalisations of logics and programming languages have been carried
out.
In addition, the formalisation of the properties of DLs also requires a
good library support for algebra (to handle the structural properties
of DLs) as well as analysis (to handle shadow-lifting), a
task for which the Mathematical Components libraries (hereafter, \mathcomp{}~\cite{mathcomp}) seem well fitted.

Our contributions in this paper are as follows:
\begin{itemize}
\item We explain how to encode known DLs in a single generic syntax
  using \coq{}, taking advantage of dependent types and building on
  known techniques for logic embedding (such as intrinsic typing).
  The formalisation is comprehensive and extensible for future use.
\item We demonstrate how to use the \mathcomp{} libraries for our
  purpose, which includes reusable lemmas that we had to newly
  develop.
\item As a result we are able to find and fix errors in the literature.
  The most prominent missing results were: soundness of \STL{} and
  missing parts of the shadow-lifting proofs, both of which appear as
  original results in this paper.
\end{itemize}
 
The paper proceeds as follows.
Sect.~\ref{sec:background} provides further background information
about property-guided training and DLs.
Sect.~\ref{sec:syntax} explains how one can define DLs in \coq{} using a generic
encoding, including a translation function producing the semantics.
Sect.~\ref{sec:logical} focuses on the formalisation of logical
properties and soundness of DLs.
In Sect.~\ref{sec:differentiability}, we demonstrate the formal verification of 
the
shadow-lifting properties of DLs.
We discuss related work and conclude
in Sect.~\ref{sec:conclusion}.
	
\section{Background}
\label{sec:background}

\subsection{Property-guided training, by means of an example}\label{sec:ex}

\paragraph*{Neural network properties}
Given a neural network $N: \Real^m \to \Real^n$, the verification property 
usually takes the form of a Hoare triple
$\forall \x \in \Real^m . \Pp(\x) \longrightarrow \Sp(\x)$,
 where $\Pp$ and~$\Sp$ can be arbitrary properties 
obtained by using variables $\x \in \Real^m$,
constants, vector, arithmetic operations, $\leq$, $=$, $\land$, $\lor$, and $\neg$. Additionally, 
$\Sp$ may contain the neural network $N$ as a function.
 
\begin{example}[Properties of neural networks]\label{ex:prop}
Given a neural network $N$ and a vector $\ve$, consider the
specification that requires that for all inputs $\x$ that are within
$\epsilon$ distance from~$\ve$, the output of $N(\x)$ should not
deviate by more than $\delta$ from $N(\ve)$:
 
 $$\forall \x. | \x - \ve|_{L_{\infty}} \leq \epsilon \Rightarrow | N(\x) - 
 N(\ve)|_{L_{\infty}} \leq \delta.$$ 
 
\noindent This  property is known as
\newterm{$\epsilon$-$\delta$-robustness}~\cite{CasadioKDKKAR22}.
It can be used to avoid misclassifying images when only a few pixels are perturbed. 
This particular example uses the \newterm{$L_{\infty}$ norm}:
$| \x - \y|_{L_{\infty}} \mydef \max_{i\in\{0,\ldots,n-1\}} ([\x]_i - [\y]_i)$,
where $[\x]_i$ stands for the $i$th element of~$\x$.

\end{example}

Unfortunately, as demonstrated by Fischer et al.~\cite{fischer2019dl2}, even most accurate neural networks 
fail even the most natural verification properties, such as $\epsilon$-$\delta$-robustness. This motivated 
the search for better ways to train the networks.  

\paragraph*{Property-guided training} Methods for property-guided
training have received considerable attention in the AI literature, as the 
survey~\cite{ijcai2022p767} shows.
We will only illustrate the method that was suggested by Fischer et
al.~\cite{fischer2019dl2}, and refer the reader to the survey for more
examples.

\begin{example}[Generating a loss function from a logical property~\cite{fischer2019dl2}]   
Recall that standard supervised learning trains a neural network $N$ with trainable parameters $\theta$
to optimise the objective $ \min_{\theta} \lossfn(\x, \y)$,
for the loss function $\losssymbol: \Real^m \times \Real^n \rightarrow \Real$.
Generally, $\losssymbol$ measures the difference between 
the network's output and the given data for each input point. Examples of 
$\losssymbol$ are cross-entropy loss or mean squared error.
But now we want to train the neural network to satisfy any arbitrary property
$\forall \x. \Pp(\x) \longrightarrow \Sp(\x)$. For this,  
we replace the above optimisation objective  with
$$\min_{\theta} \left[\max_{\x \in \mathbb{H}_{\Pp(\x)}} \lossfn_{\Sp}(\x)\right]$$
where $ \mathbb{H}_{\Pp(\x)} \subseteq \Real^m$ refines the type $\Real^m$ to a subset for which the property $\Pp$ holds, 
  and $\lossfn_{\mathcal{S}}: \Real^m  \rightarrow \Real$ is obtained by applying a suitable
\newterm{interpretation function} for $\Sp$.

We omit the exact details of how such optimisation algorithms are
defined: they are known and can be found in a suitable machine
learning tutorial, for example~\cite{KM18}.
Intuitively, the optimisation algorithm will search for $\x \in \mathbb{H}_{\Pp(\x)}$ such that $\x$ 
maximises the loss $\lossfn_{\Sp}(\x)$, 
in order to train the neural network parameters $\theta$ to minimise that loss. 
Concretely, if the property is $\epsilon$-$\delta$-robustness, it will look for the worst perturbation of $\ve$
that violates the property, and will optimise the neural network to classify that bad example correctly. 
\end{example}

\paragraph*{Differentiable logics for loss functions}  In the above
example, we did not explain how to define the interpretation function
$\lossfn_{\Sp}$ for an arbitrary property $\Sp$; we need differentiable
logics for that purpose.
Fischer et al.~\cite{fischer2019dl2} proposed one such interpretation
function---called \newterm{the differentiable logic \DLtwo{}},
standing for ``Deep Learning with Differentiable Logics''.


\begin{example}[Loss functions from properties in a fuzzy logic]\label{ex:loss}

Taking the properties from Example~\ref{ex:prop}, by the Fischer et al.\ method
we must be able to interpret the right-hand side of the implication, i.e.,
$| N(\x) - N(\ve)|_{L_{\infty}} \leq \delta$, 
given concrete values for $\epsilon$, $\delta$, 
a concrete vector~$\ve$, neural network~$N$, and a suitable definition of the 
$L_{\infty}$ norm.
For example, interpretation for our property in \STL{}~\cite{varnai} is given by
$
 \tempty{| N(\x) - N(\ve)|_{L_{\infty}} \leq \delta}_{\STL} = $
$\delta - |\tempty{N(\x)} - \tempty{N(\ve)}|_{L_{\infty}}$.
On the left-hand side, the $L_{\infty}$ distance between vectors as well as~$N$ are defined
in the syntax of \STL{}; on the right-hand side, they are given by real vector
arithmetic operations. Example~\ref{ex:ldlcoq} will make the relation between syntax and interpretation clear. 
The obtained function can be used directly for training neural networks.
\end{example}

We next consider our choices of DLs in details.
 
\subsection{Differentiable logics}
 
Ślusarz et al.~\cite{ldl} suggest a common syntax for all DLs, calling it
the \newterm{logic of differentiable logics (LDL)}, and 
subsequently obtain different DLs via different interpretation functions. 
In the following, we summarize the syntactic and semantic features of DLs following this formulation; 
minor modifications will be discussed as we introduce them.

\begin{figure}
\begin{subfigure}[t]{0.9\textwidth}
\small
\text{type} $\ni t$ ::= \BoolType\ | \FinType{n}   for $n \in \Nat$
|  \RealType\ |  \VecType{n}\  | 
 \FunType{n}{m}  for $n,m \in \Nat$ \\
\end{subfigure}
\begin{subfigure}{0.5\textwidth}
	\small
	\begin{tabular}{rlcl}
		\text{exprInd} & $\ni i$ & ::= & $i \in \Nat$ \\ 
		\text{exprR} & $\ni r, r_1, r_2$ & ::= &  $r \in \Real$  | $[v]_i$  \\
		\text{exprFun} & $\ni f$ & ::= & $f \in \mathbb R^n \to \mathbb R^m$ \\
		\text{exprVec} & $\ni v$ & ::= & $v \in R^n  $ | $f\ v$ \\
	\end{tabular}
\end{subfigure}
\begin{subfigure}{0.35\textwidth}
	\small
	$$
	\begin{array}{rlcl}
		\text{exprB} &\ni p, p_0, \ldots, p_n & ::= & \True{}\ |\ \False{} \\
		& & | & r_1 \leq r_2 \\
		& & | & r_1 \neq r_2 \\
		& & | & \naryConj (p_0, \ldots, p_M) \\
		& & | & \naryDisj (p_0, \ldots, p_M)  \\
		& & | & \neg p \\
	\end{array}
	$$
\end{subfigure}
\caption{Types and expressions of LDL.}
\label{fig:syntax-types-math}
\end{figure}

\paragraph*{LDL syntax}
LDL's syntax consists of types and expressions (Fig.~\ref{fig:syntax-types-math}). 
Types are given by booleans, reals, vectors, indices, and a function type  \FunType{n}{m}; 
expressions are given by real numbers, vectors, vector indices, lookup operations,
and functions that take real vectors as inputs.
Formulae are formed either via applying predicates $\leq, =$ to real expressions,
by boolean values, or using logical connectives $\lor, \land, \neg$.
Because \STL{} by Varnai et al.~\cite{varnai} lacks associativity,
conjunction and disjunction are defined as $n$-ary connectives 
so that they are the same for all DLs.
Further, implication is not present in the syntax: that is due 
to the $n$-ary nature of the other connectives, which do not always 
allow for the implication of classical logic. Any DL with associative 
conjunction and disjunction
will admit implication $b_1 \impl b_2$ to be defined as $\neg b_1 
\vee b_2$.
 
We forgo the originally included quantifiers, lambda, and 
let expressions to obtain a simpler core language in which
the three properties of interest---soundness, compositionality, and differentiability---can be studied. 
 
\paragraph*{Obtaining DLs via interpretation functions}
To define a DL, one defines an interpretation function $\tempty{\cdot}_{\DL}$ that, given an expression in LDL, 
returns a function on real numbers.  We introduce all DLs in a generic way and use the meta-notation
$\tempty{\cdot}_{\DL}$, to refer to a range of interpretation functions, with
$\DL \in \{\Godel, \Luka, \Yager, \product, \DLtwo, \STL\}$.
The boolean interpretation function $\tbool{\cdot}$ is the obvious 
logical interpretation of boolean formulas, which will be useful for proving 
soundness later.
 
Table~\ref{tab:semantics-math} shows the interpretation of all
DLs. First are the four DLs based on well-known fuzzy logics: \Godel,
\Luka~\cite{lukasiewicz1920three}, \Yager{}, and \product{}~\cite{van2022analyzing}.
All fuzzy logics have the interpretation domain of $[0,1] \subset \Real$.  Other logics have
different domains: \DLtwo~\cite{fischer2019dl2} has the interpretation
domain $(- \infty , 0] $, and \STL~\cite{varnai} the domain $(-\infty, +\infty)$.

The binary predicates $\leq$ and $=$ are defined in a way that ensures that they are interpreted within 
the chosen real interval for the  given DL.  The definitions of logical 
connectives $\naryDisj$, $\naryConj$, and $\neg$ are taken directly 
from the related papers that define the given DLs. Note that we reformulate 
$\naryDisj$ and $\naryConj$ for all DLs as $n$-ary connectives, 
however, only \STL{} had $n$-ary connectives originally. 

\newcommand{\topline}{\arrayrulecolor{black}\specialrule{0.1em}{\abovetopsep}{0pt}%
	\arrayrulecolor{lightgray}\specialrule{\belowrulesep}{0pt}{0pt}%
	\arrayrulecolor{black}
}

\newcommand{\midline}{\arrayrulecolor{lightgray}\specialrule{\aboverulesep}{-1pt}{0pt}%
	\arrayrulecolor{black}\specialrule{\lightrulewidth}{0pt}{0pt}%
	\arrayrulecolor{white}\specialrule{\belowrulesep}{0pt}{0pt}%
	\arrayrulecolor{black}
}

\begin{table}[t]
\footnotesize

\newcommand\TL{\rule{0pt}{3.3ex}}
\newcommand\BL{\rule[-2.1ex]{0pt}{0pt}}
\newcommand\TY{\rule{0pt}{4ex}}
\newcommand\BY{\rule[-2.8ex]{0pt}{0pt}}
\newcommand\BP{\rule[-1.5ex]{0pt}{0pt}}
\newcommand\BDLtwo{\rule[-1.5ex]{0pt}{0pt}}

{\renewcommand{\arraystretch}{1.5}
\begin{tabularx}{\textwidth}{|p{5.2em}|l|l|l|}
\hline
 \rowcolor{lightgray}            & $ \tempty{ \naryConj s } $ & $ \tempty{ \naryDisj s } $ & $ \tempty{ \lnot e }$ \\
\hline
\text{G\"odel}      & $ \min\tGodel{s} $ & $ \max\ \tGodel{s} $ & $ 1-\tGodel{e}$ \\
\hline
\text{Łukasiewicz}  & $ \max\left[\sum_{a \in \tlukasiewicz{s}} a - |s|+1, 0\right] $ \TL \BL & $ \min\left[\sum_{a \in \tlukasiewicz{s}} a, 1\right] $ & $ 1-\tlukasiewicz{e}$ \\
\hline
\text{Yager}        & $ \max\left[1 - \left(\sum_{a \in \tyager{s}} (1-a)^p\right)^{1/p}, 0\right] $ \TY \BY & $ \min\left[\left(\sum_{a \in \tyager{s}} a^p\right)^{1/p}, 1\right] $ & $ 1-\tyager{e}$ \\
\hline
\text{product}      & $ \prod_{a \in \tproduct{s}} a $ \BP & $ \fold\ (\lam{x\ y}\ x+y-xy)\ 0\ \tproduct{s} $ & $ 1-\tproduct{e}$ \\
\hline
\text{\DLtwo}       & $ \sum_{a \in \tDLtwo{s}} a $ \BDLtwo & $ (-1)^{|s|+1}  \cdot \prod_{a \in \tDLtwo{s}} a $ & $ \text{undefined${}^\dag$} $ \\
\hline
\text{\STL}         & $ \mathit{and}_S\ \tSTL{s} $ & $ \mathit{or}_S\ \tSTL{s} $ & $ {-}\tSTL{e} $\\
\hline
\text{Bool}         & $\naryConj \tbool{s}$ & $\naryDisj \tbool{s}$ & $\neg \tbool{e} $ \\
\hline
\end{tabularx}
}

\newcommand\BF{\rule[-4ex]{0pt}{0pt}}

{\renewcommand{\arraystretch}{1.3}
\begin{tabularx}{\textwidth}{|p{5.2em}|p{12.6em}|l|l|l|}
\hline
\rowcolor{lightgray}                        & $ \tempty{ e_1 = e_2 } $ & $ \tempty{ e_1 \le e_2 } $ & $ \tempty{ \True } $ & $ \tempty{ \False }$ \\
\hline
\text{fuzzy}        & $\begin{array}{l}
                         \text{if }{\tempty{e_1} = -\tempty{e_2}}\\
    		     \text{then }{\tempty{e_1} = \tempty{e_2}}\\
                         \text{else }{\max\left[1-\left|{\tempty{e_1} - \tempty{e_2} \over \tempty{e_1} + \tempty{e_2}}\right|, 0\right]}
                       \end{array} $ &
                      $\begin{array}{l}
                         \text{if }{\tempty{e_1} = -\tempty{e_2}}\\
                         \text{then }{\tempty{e_1} \le \tempty{e_2}}\\
                         \text{else}\\
                         {\max\left[1-\max\left[{\tempty{e_1} - \tempty{e_2} \over \tempty{e_1} + \tempty{e_2}}, 0\right], 0\right]}
                       \end{array} $ & $ 1 $ & $ 0 $\\
\hline
\text{\DLtwo}       & $ -|\tDLtwo{e_2} - \tDLtwo{e_1}| $ & $ - \max\left[\tDLtwo{e_1}-\tDLtwo{e_2}, 0\right] $ & $ 0 $ & $ -\infty $ \\
\hline
\text{\STL}         & $ -|\tSTL{e_2} - \tSTL{e_1}| $ & $ \tSTL{e_2} - \tSTL{e_1} $ & $ +\infty $ & $ -\infty $ \\
\hline
\text{Bool} & $  \tbool{e_1} = \tbool{e_2}  $ & $ \tbool{e_1} \le \tbool{e_2} $ & $  \True $ & $  \False $ \\
\hline
\end{tabularx}
}

\begin{tabular}{lcc}
$
and_S\ [a_1,\ldots,a_M]  =
\begin{cases}
  \dfrac{\sum_i a_{\min} e^{\tilde{a_i}} 
  e^{\nu 
  \tilde{a_i}}}{\sum_i e^{\nu 
  \tilde{a_i}}} & 
                      \text{if}\ 
                      a_{\min} < 0 \\
  \dfrac{\sum_i a_i e^{-\nu 
  \tilde{a_i}}}{\sum_i e^{-\nu 
  \tilde{a_i}}} & \text{if}\ 
                      a_{\min} > 0 \\
  0 & \text{if}\ a_{\min} = 0 \\
\end{cases}
$ 
\quad where \quad
$
\begin{array}{rcl}
  \nu &\in& \Real^+ \text{ (constant)} \\
  a_{\min} &=& \min\ [a_1, \ldots, a_M] \\
  \tilde{a_i} &=& \dfrac{a_i - a_{\min}}{a_{\min}}		
\end{array}
$ \\[2em]
$or_S$ is analogous to $and_S$ & &
\end{tabular}

\caption{Interpretation function $\tempty{\cdot}_{\DL}$, which extends naturally
to sequences of expressions as well as interpretation function $\tbool{\cdot}$,
which is a structural interpretation of boolean formulas.\\
{\footnotesize $\dag$: Negation is undefined in the sense that it is not defined
as a structural transformation of the syntax of formulas. It is implemented at
the level of atomic comparison only, and negation on composite formulas is provided
as syntactic sugar~\cite[Sect.~3]{fischer2019dl2}. For example, consider
$\tempty{3 = 3}_{\DLtwo} = 0$. In $\DLtwo$, $\tempty{\neg (3 = 3)}_{\DLtwo}$ is not
defined in term of  $\tempty{3 = 3}_{\DLtwo}=0$, but $\DLtwo$ provides an
interpretation for the symbol $\neq$ separately.}}
\label{tab:semantics-math}
\end{table}

\subsection{Properties of DLs}
\label{sec:propertiesDLs}

\paragraph*{Soundness}
There is no consensus in the DL literature on how or whether to state soundness:
for example, \STL{} came without any soundness statement.
For the sake of generic formalisation of all DLs, we propose the following definition of soundness,
which generalises soundness as defined in \DLtwo{} and fuzzy 
logics~\cite{fischer2019dl2,van2022analyzing}. 

\begin{definition}[Soundness]
	\label{def:sound}
Given a DL, an expression $e$, and a boolean value $b \in 
\{\True, \False\}$, the DL is sound if 
\begin{equation*}
\tempty{e}_{\DL{}} = \tempty{b}_{\DL{}} \implies \tbool{e} = b
\end{equation*}
\end{definition}

Note that not all DLs are sound. For example,  one of the oldest fuzzy logics 
by  
Łukasiewicz~\cite{lukasiewicz1920three} is known to be unsound. 
Table~\ref{tab:properties} summarises all known soundness results. Note that 
prior to this paper, soundness of \STL\ was not known. Here, we obtain the
result with some restrictions, see Sect.~\ref{sec:logical}.

\paragraph*{Compositionality} 
We define idempotence, associativity, and 
commutativity of interpretation functions for $\naryConj$ and analogously 
$\naryDisj$:

\begin{definition}[Commutativity, idempotence, and associativity of  $\naryConj$]
Given a DL, the interpretation function of conjunction is 
\newterm{commutative} if for any permutation $\pi$ of the integers $i \in \{1, \ldots, M\}$
we have
\begin{equation*}
	\tdlbig{\naryConj \left(\pval_0, \ldots, \pval_M\right)} =
        \tdlbig{\naryConj \left(\pval_{\pi(0)}, \ldots, \pval_{\pi(M)}\right)}.
\end{equation*}
It is \newterm{idempotent} and \newterm{associative} if we have
\begin{align*} 
\tdlbig{\naryConj(\pval, \ldots, \pval)} &= \tdl{\pval}, \\
\tdlbig{\naryConj\left(\naryConj(\pval_0, \pval_1), \pval_2\right)} &= 
\tdlbig{\naryConj\left(\pval_0, \naryConj(\pval_1, \pval_2)\right)}.
\end{align*} 
\end{definition}

Table~\ref{tab:properties} shows which DLs satisfy which logical properties.
Finally, as already illustrated in Sect.~\ref{sec:ex}, negation can be problematic in some DLs.
For example, \DLtwo{} does not give a direct interpretation for negation, as its domain is asymmetric.
We will see in Sect.~\ref{sec:logical} that negation also causes problems with the soundness of \STL{}. 

\paragraph*{Differentiability}
Varnai et al.~\cite{varnai} introduce three properties in this category: weak 
smoothness, scale-invariance, and shadow-lifting.
The latter is the most important as it accounts for gradual improvement in 
training.
We only consider shadow-lifting here as it is the most complex of those 
properties and leave the remaining properties to future work.

\begin{definition}[Shadow-lifting property~\cite{varnai}]
\label{def:shadow}
The DL satisfies the \newterm{shadow-lifting} property if, for any
$ \tdl{\pval} \neq 0$:
\begin{equation*}
	\left. \dfrac{\partial \tdlbig{\naryConj(\pval_0, \ldots, \pval_i, \ldots, 
			\pval_M)}}{\partial 
		\tdl{\pval_i}}\right\rvert_{\pval_j = \pval \text{ where } i 
		\neq j} >0
\end{equation*}
holds for all $0 \leq i \leq M$,
where $ \partial $ denotes partial differentiation.
\end{definition}

\noindent Notice that classical conjunction does not satisfy the
property of shadow-lifting: no matter how ``true'' the value of $p_2$ is, if $p_1$ is
false, then $p_1 \land p_2$ will remain false. Likewise, all DLs that use $\min$ or
$\max$ to define conjunction will fail shadow-lifting.

Shadow-lifting was originally defined for conjunction only, as \STL{} had no 
disjunction. 
In our formalisation, we could, in principle, extend shadow-lifting to disjunction.
However, we left this incremental extension for future work.    

\begin{table}[t]
{\footnotesize
	\begin{tabular}{|p{0.15\textwidth}|c|c|c|c|c|c|}
		\hline 
		\textbf{Properties:} &  Negation & Idempotence & Commutat. & Associativ. & Soundness & Shadow-lifting  \\ 
		\hline 

		   \Godel & yes & yes & yes & yes  & yes & no \\
		\hline 

		\Luka & yes &no & yes & yes  & no& no \\
		\hline 
		\Yager  & yes  & no & yes &yes & no & no 
		\\
		\hline 
		\product  &yes & no &yes &yes & yes & \cellcolor{orange} yes \\
		\hline

		\DLtwo  & no &  no &  \cellcolor{orange} yes & 
		\cellcolor{orange} yes & \cellcolor{yellow} yes$^\dagger$  & 
		 \cellcolor{orange} yes
		\\
		\hline 			
		\STL   & yes &yes &yes & no & 
		\cellcolor{orange} yes$^\dagger$ & \cellcolor{yellow} yes \\
		\hline 
	\end{tabular}}
	
			\caption{Properties of the different DLs formalised in this paper \cite{github}. We distinguish previously known proofs that we mechanise from
			previously known results published with incomplete or
				semi-formal proofs (\colorbox{yellow}{yellow}) and new results (\colorbox{orange}{orange}).\\
 {\footnotesize $\dag$: For \DLtwo{} and \STL{}, we prove soundness of the
  negation-free fragment of LDL; negation
 is undefined for \DLtwo, and \STL{} is not sound for the
 full fragment.}}
\label{tab:properties}
\setlength{\belowcaptionskip}{-15pt}
\vspace*{-1em}
\end{table}

\paragraph*{Summary of results}
Table~\ref{tab:properties} summarises all properties covered in our \coq{} 
formalisation and highlights the ones for which we provide 
original proofs.
In our development, we provided several missing results, most prominently,
the soundness of \STL{} and missing parts of the shadow-lifting proofs. 
Note that the formalisation further revealed some errors:
\begin{example}[Discrepancies in pen and paper proofs]\label{ex:errors}
While doing the \coq{}  formalisation, we found two sources of errors in our pen and paper proofs \cite{ldl} as well as in~\cite{varnai}.
Firstly, the work in \cite{ldl} concerned completing results of Table~\ref{tab:properties} in the uniform notation of LDL. Many proofs were analogous and it was easy to overlook the rare cases when the proof could not be completed by analogy with existing proofs. For example, we tried to prove the soundness of \Yager{} by analogy with other fuzzy logics overlooking that \Yager{} is not sound. Indeed \Yager{} is a generalisation of \Luka{} logic which in itself is not sound.
The \coq{} formalisation revealed all such errors.

The second source of errors came from extension of fuzzy logics with comparison operators. The interpretation of comparisons needed to be scaled between $0$ and $1$, and it seemed obvious that the scaling was done correctly. Therefore we did not provide any proofs concerning the interval properties of these operations. In contrast the soundness proofs in \coq{} required us to prove that the fuzzy interpretation functions always return values within the interval~$[0,1]$. 

Thirdly, while a sketch of the shadow-lifting proof was provided 
in~\cite{varnai}, it was incomplete and did not mention either the existence of 
other cases as well as the reliance of the proof on L'H\^opital's rule.
\end{example}
No DL satisfies all desirable properties---for example, \Godel{}
is sound, idempotent, associative, and commutative, but it is not
shadow-lifting. On the other hand, \Luka{} is not sound, \STL{} not
associative, and while \DLtwo{} is sound and shadow-lifting, it fails
idempotence, and its negation is not compositional because it is not structural.
Varnai et al.~\cite{varnai} have proven that it is impossible for any DL to be
idempotent, associative, and shadow-lifting at the same time.

When one has to make a choice of a DL, different considerations may influence that choice.
Soundness and shadow-lifting are strictly desirable, thus \Godel{}, \Luka{} 
and \Yager{} are probably less desirable than the rest, even if some of them have nice logical properties.  
However, given soundness and shadow-lifting, the choice between logical properties is less clear.
For example, one can imagine a scenario when the specification
language avoids negation,
and in a style of substructural logics,
 treats differently $p$ and $p \land p$ and thus sacrifices idempotence; in this case, 
\DLtwo{} may
provide an ideal translation function.

\section{An encoding of DLs in \coq}
\label{sec:syntax}

As discussed, LDL aims at defining all DLs in a generic and extendable
way, using uniform syntactic conventions.  In this section, we start
by highlighting the generic features of our formalisation.

\subsection{Encoding of the syntax of types and expressions}

The encoding of the LDL types is the matter of declaring the following
inductive type:
\begin{minted}[fontsize=\footnotesize]{ssr}
Inductive flag := def | undef.

Inductive ldl_type :=
  Bool_T of flag | Index_T of nat | Real_T | Vector_T of nat | Fun_T of nat & nat.
\end{minted}
This corresponds to the informal syntax of
Fig.~\ref{fig:syntax-types-math} with the difference that we refine the Boolean type with a (\coqin{flag}) to
signify whether negation is defined in the logic. 

As for LDL expressions, their encoding is displayed in
Fig.~\ref{fig:syntax-coq}.  It is an inductive type indexed by
\coqin{ldl_type} so that the resulting syntax is intrinsically-typed:
one cannot write ill-typed expressions.
The \coq{} inductive type matches the informal syntax already
explained in Fig.~\ref{fig:syntax-types-math}.
Real expressions (line \ref{line:ldlreal}) use a type \coqin{R} of
type \coqin{realType} coming from \analysis{}~\cite{analysis} that represents real
numbers.
Boolean expressions (line \ref{line:ldlbool}) use the native \coq{}
type \coqin{bool}.
Indices embed an \newterm{ordinal\/} from \mathcomp\ (line~\ref{line:ldlidx}).
More specifically, \coqin{'I_n} is the type of natural number smaller than \coqin{n}.
Similarly, vectors just reflect \mathcomp{}
tuples (line~\ref{line:ldlvec}).
For defining $n$-ary connectives \coqin{ldl_and} and \coqin{ldl_or},
we use polymorphic lists (of type \coqin{seq}).
Yet, to ease reading, we will use notation such as \coqin{a `/\ b} to denote binary conjunction
in the following.
For a generic definition of the syntax, we need to allow for the case
of DLs in which negation is not defined (in fact, \DLtwo).
The additional argument \coqin{Bool_T_def} in the constructor
\coqin{ldl_not} (line \ref{line:ldlnot}) signifies that the negation is defined.
%
The constructor \coqin{ldl_cmp} (line \ref{line:ldlcmp}) is for binary
comparison operators over the real numbers; hereafter, we will use
notations such as \coqin{`<=} for the comparison corresponding to
$\leq$ instead of ``\coqin{ldl_cmp cmp_le}'' to ease reading.
As for the last constructors, they are respectively for functions,
their application, and lookups, as per Fig.~\ref{fig:syntax-types-math}.

\begin{figure}[h]
\begin{minted}[fontsize=\footnotesize,numbers=left,xleftmargin=1.5em,escapeinside=88]{ssr}
Definition Bool_T_undef := Bool_T undef.
Definition Bool_T_def := Bool_T def.
Inductive comparison := cmp_le | cmp_eq.

Inductive expr : ldl_type -> Type :=
 | ldl_real   : R -> expr Real_T 8\label{line:ldlreal}8
 | ldl_bool   : forall p, bool -> expr (Bool_T p) 8\label{line:ldlbool}8
 | ldl_idx    : forall n, 'I_n -> expr (Index_T n) 8\label{line:ldlidx}8
 | ldl_vec    : forall n, n.-tuple R -> expr (Vector_T n) 8\label{line:ldlvec}8
 | ldl_and    : forall x, seq (expr (Bool_T x)) -> expr (Bool_T x)
 | ldl_or     : forall x, seq (expr (Bool_T x)) -> expr (Bool_T x)
 | ldl_not    : expr Bool_T def -> expr Bool_T def 8\label{line:ldlnot}8
 | ldl_cmp    : forall x, comparison -> expr Real_T -> expr Real_T -> expr (Bool_T x) 8\label{line:ldlcmp}8
 | ldl_fun    : forall n m, (n.-tuple R -> m.-tuple R) -> expr (Fun_T n m)
 | ldl_app    : forall n m, expr (Fun_T n m) -> expr (Vector_T n) -> expr (Vector_T m)
 | ldl_lookup : forall n, expr (Vector_T n) -> expr (Index_T n) -> expr Real_T.
\end{minted}
\caption{LDL syntax in \coq.}
\label{fig:syntax-coq}
\end{figure}

\subsection{Encoding of the interpretation function}

We now proceed to the translation function that
interprets the syntax.
Types are mapped to their obvious semantics:
\begin{minted}[fontsize=\footnotesize,numbers=left,xleftmargin=1.5em,escapeinside=88]{ssr}
Definition type_translation (t : ldl_type) : Type :=
match t with
| Bool_T x => R 8\label{line:transbool}8
| Real_T => R
| Vector_T n => n.-tuple R
| Index_T n => 'I_n
| Fun_T n m => n.-tuple R -> m.-tuple R
end.
\end{minted}
In particular, booleans are mapped to \coqin{R} of type \coqin{realType}, the type of real numbers.
This translation accommodates the many interpretations of the DLs: the domain
$[0,1]$ for fuzzy logic, $(-\infty, 0]$ for \DLtwo{}, and $(-\infty, +\infty)$ for \STL{}.
For \DLtwo{} and \STL{}, we also provide an alternative translation function \coqin{ereal_type_translation}
that maps booleans to real numbers extended with $-\infty$ and $+\infty$, i.e., the type \coqin{\bar R} of extended real numbers
as provided by \analysis.
We use an invariant to restrict the range of the interpretation function accordingly.

Each logic requires a separate interpretation function (as explained
in Table~\ref{tab:semantics-math}).  Here, we only show an excerpt of
the translation function for \STL{}, namely the cases for conjunction
(constructor \coqin{ldl_and}), negation (notation \coqin{`~}), and comparison
(notation \coqin{`<=}) of \STL{} in Fig.~\ref{fig:semantics-coq}.

\begin{figure}[h]
\begin{minted}[fontsize=\footnotesize,xleftmargin=1.5em,escapeinside=88]{ssr}
Fixpoint stl_translation {t} (e : expr t) : type_translation t :=
  match e in expr t return type_translation t with
  | ldl_and _ (e0 :: s) => let A     := map stl_translation s in       
                           let a0    := stl_translation e0 in          
                           let a_min := \big[minr/a0]_(i <- A) i in    
                           if a_min < 0 then stl_and_lt0 (a0 :: A) else
                           if a_min > 0 then stl_and_gt0 (a0 :: A) else
                           0                                           
  | `~ E1               => - {[ E1 ]}
  | E1 `<= E2           => {[ E2 ]} - {[ E1 ]}
  ... (* see 8\cite{github}8 for omitted connectives *)
  end where "{[ e ]}" := (stl_translation e).
\end{minted}
\setlength{\belowcaptionskip}{-5pt}
\caption{Excerpt of the semantics of \STL{}: conjunction, negation, and comparison. \\
{\footnotesize (See Fig.~\ref{fig:semantics-lt0-gt0} for intermediate definitions \coqin{stl_and_lt0} and \coqin{stl_and_gt0}.)}}
\label{fig:semantics-coq}
\end{figure}

The case for conjunction of \STL{} is the most complex in our
formalisation because dealing formally with it requires the theories
of exponentiation (\coqin{expR}), big sums (\coqin{\sum}), inverses
(\coqin{^-1}), and minima (\coqin{minr}).
To reduce the clutter, we define the cases for $a_{\min} > 0$ and $a_{\min} < 0$ separately
as \coqin{stl_and_gt0} and \coqin{stl_and_lt0} reproduced in Fig.~\ref{fig:semantics-lt0-gt0}.
This will allow us to state intermediate lemmas about sub-expressions.

\begin{figure}[h]
\begin{subfigure}[]{0.66\textwidth}
\begin{minted}[fontsize=\footnotesize,xleftmargin=1.5em,escapeinside=88]{ssr}
Definition sumR a := \sum_(i <- a) i.
Definition min_dev {R : realType} (x:R) (a:seq R) : R :=
  let r := \big[minr/x]_(i <- a) i in (x - r) * r^-1.
		
Definition stl_and_gt0 (a : seq R) :=
  sumR (map (fun x => x * expR(-nu * min_dev x a)) a) *
  (sumR (map (fun x => expR(-nu * min_dev x a)) a))^-1.
  
Definition stl_and_lt0 (a : seq R) :=
  sumR (map (fun x => (\big[minr/x]_(i <- a) i) *
    expR (min_dev x a) * expR(nu * min_dev x a)) a) *
  (sumR (map (fun x => expR(nu * min_dev x a)) a))^-1.
\end{minted}
\end{subfigure}
\;
\vrule
\;
\begin{subfigure}[]{0.3\textwidth}
\footnotesize
$
\begin{array}{l}
a_{\min} = \min\ [a_1, \ldots, a_M] \\
\tilde{a_i} = \dfrac{a_i - a_{\min}}{a_{\min}}\\
\\
\nu \in \Real^+
\quad\quad\quad\;\;\;\text{ (constant)} \\
\dfrac{\sum_i a_i e^{-\nu \tilde{a_i}}}{\sum_i e^{-\nu \tilde{a_i}}}
\quad\quad\text{(case }a_{\min} > 0\text{)} \\
\\
\dfrac{\sum_i a_{\min} e^{\tilde{a_i}} 
	e^{\nu 
		\tilde{a_i}}}{\sum_i e^{\nu 
		\tilde{a_i}}}  
\;\text{(case }a_{\min} < 0\text{)} \\
\end{array}
$
\end{subfigure}
\setlength{\belowcaptionskip}{5pt}
\caption{Intermediate definitions to define the conjunction of \STL{}.\\
{\footnotesize (The right subfigure reproduces part of Table~\ref{tab:semantics-math} for reading convenience.)}}
\label{fig:semantics-lt0-gt0}
\end{figure}

As a first application of our encoding, we can already formalise our
running example, with the advantage of encoding it once for all DLs:
\begin{example}\label{ex:ldlcoq}
Taking the interpretation task of Example~\ref{ex:loss}, 
we first give a suitable definition for $L_{\infty}$ norm (\coqin{ldl_norm_infty}) 
and vector subtraction (\coqin{ldl_vec_sub }).
One only needs to call the defined interpretation function 
to encode the loss function of Example~\ref{ex:loss}:

\begin{minted}[fontsize=\footnotesize]{ssr}
Context (eps delta : @expr R Real_T) (f : @expr R (Fun_T n.+1 m.+1)) 
        (v : @expr R (Vector_T n.+1)) (x : @expr R (Vector_T n.+1)).

Definition eps_delta_robust : expr Bool_T_undef :=
  ldl_lookup
    (ldl_app (ldl_norm_infty m) (ldl_vec_sub (ldl_app f x) (ldl_app f v)))
    (ldl_idx ord0) `<= delta.
\end{minted}
\end{example}

\section{Logical properties and soundness of DLs}
\label{sec:logical}

\subsection{Logical properties of DLs}

The logical properties of DLs are idempotence, commutativity, and
associativity.  Not all DLs have the same properties as we saw earlier
(Table~\ref{tab:properties}).  Proving the logical properties
essentially amounts to showing that the semantic interpretation does
have them. For example, the conjunction of \DLtwo{} being interpreted
as addition on reals inherits its associativity directly from the properties
of real numbers, and as a consequence, its proofs are one-liners, e.g.:
\begin{minted}[fontsize=\footnotesize]{ssr}
Lemma dl2_andA (e1 e2 e3 : expr Bool_T_undef) :
  [[ e1 `/\ (e2 `/\ e3) ]]_dl2 = [[ (e1 `/\ e2) `/\ e3 ]]_dl2.
Proof. by rewrite /=/sumR ?big_cons ?big_nil !addr0 addrA. Qed.
\end{minted} 

In contrast, for \Yager{} and \STL{}, the proofs are more demanding.  For
example, the associativity for \Yager, though stated analogously,
\begin{minted}[fontsize=\footnotesize]{ssr}
Theorem Yager_andA (e1 e2 e3 : expr Bool_T_def) :
   0 < p ->  [[ (e1 `/\ e2) `/\ e3]]_Yager = [[ e1 `/\ (e2 `/\ e3) ]]_Yager.
\end{minted} 
consists of about 100 lines of code. This is because in this case, the
interpretation relies on the power function of \analysis{} whose
properties are more technical. Yet, we could put the automatic tactics
available with \mathcomp{} such as \coqin{lra}~\cite{sakaguchi2022itp,algebratactics} to good use.

\subsection{Soundness of DLs}
\label{sec:sound}

We now address the topic of formalising the soundness results of
Table~\ref{tab:properties}.

\subsubsection*{Soundness for closed interval DLs}
For fuzzy DLs and, more generally, closed interval DLs there is a clear consensus on 
how to define soundness: we generalised it in Definition~\ref{def:sound}.
It boils down to taking the least and greatest elements in the given real interval as 
interpretations for $\False$ and $\True$, respectively.
In \coq, the statement of soundness for \Godel{} and \product{} is as follows:
\begin{minted}[fontsize=\footnotesize]{ssr}
Lemma soundness (e : expr Bool_T_def) b :
  [[ e ]]_ l = [[ ldl_bool _ b ]]_ l -> [[ e ]]_B = b.
\end{minted}
This is a direct paraphrase of the pencil-and-paper 
Definition~\ref{def:sound}. The proofs proceed by induction 
and require inversion lemmas, which we will discuss later in this section. 

\subsubsection*{Soundness for DLs with open intervals}
When a DL's domain of interpretation is given by an open interval, which is the 
case for  \DLtwo{} and \STL{},
there is no clear consensus in the literature on defining or proving soundness.
We will illustrate the problems that arise using \STL{} and following previous work~\cite{ldl}.
The first question is how to state soundness. The easiest choice  is to simply add $-\infty$ and $+\infty$ as 
constants to the domain, and keep the soundness statement of Definition~\ref{def:sound}.
However, because no formula in the language evaluates to $-\infty$ or $+\infty$,
such a soundness proof is vacuous.  Note that Definition~\ref{def:sound} did not 
cause this problem for fuzzy DLs because there were formulae in the language that 
evaluated to bottom and top values: take for example $\tproduct{3 = 3} = 1$.

Alternatively, one may keep the open interval intact and simply
re-define soundness in terms of intervals: if the interpretation of
the formula $e$ is greater or equal to $0$, then $\tbool{e} = \True$
else $\tbool{e}=\False$.
However, this solution triggers a different problem: negation is no
longer sound. Indeed, if $\tempty{3 = 3}_{\STL} = 0$ means the formula
is true, then the same can be said about
$\tempty{\neg (3 = 3)}_{\STL} = 0$.

One could think of a solution excluding $0$ from the interval $(-\infty, +\infty)$ altogether, but that 
complicates the interpretation of comparisons and creates a point in the interval 
at which the resulting function is not differentiable, which damages shadow-lifting.

\subsubsection*{\coq{} formalisation for logics with open intervals}

For the reasons explained above, we remove negation from \STL{} and use intervals to
define the truth:
\begin{minted}[fontsize=\footnotesize]{ssr}
Definition is_stl b (x : R) := if b then x >= 0 else x < 0.
\end{minted}
This results in the following soundness statement:
\begin{minted}[fontsize=\footnotesize]{ssr}
Lemma stl_soundness (e : expr Bool_T_undef) b :
  is_stl b (nu.-[[ e ]]_stl) -> [[ e ]]_B = b.
\end{minted}
The flag \coqin{Bool_T_undef} in \coqin{(e : expr Bool_T_undef)} signifies that
the proof omits the case that uses negation.
We write \coqin{nu.-[[ e ]]_stl} as notation for interpretation of
\STL{} (and similar notation for the remaining DLs).

The soundness proof proceeds by induction on the structure of the
interpretation function.  Because of the extensive use of dependent
types in our formalisation we need a custom dependent induction
principle.
The most interesting cases are those for conjunction and disjunction,
which need special inversion lemmas.  Here is one example:
\begin{minted}[fontsize=\footnotesize]{ssr}
Lemma stl_nary_inversion_andE1 (s : seq (expr Bool_T_undef)) :
  is_stl true (nu.-[[ ldl_and s ]]_stl) ->
    forall i, i < size s -> is_stl true (nu.-[[ nth (ldl_bool pos false) s i ]]_stl).
\end{minted}

Our formalisation faced a minor technical problem: if we are to 
comply with the generic DL syntax defined in Fig.~\ref{fig:syntax-types-math},
we need to interpret constants $\True$ and $\False$ present in the language.  
We therefore propose two alternative interpretations for \DLtwo{} and
\STL{}: one that works on extended reals (with added constants $-\infty$, $+\infty$) 
and maps $\True$ and $\False$ to the top and bottom elements of the 
respective domains, and one that
resolves this discrepancy by choosing arbitrary interpretations for
$\True$ and $\False$ that satisfy all the properties of interest for our
study.
In the latter case, for \DLtwo{} we choose to interpret $\True$ as $0$ and
$\False$ as $-1$, and for \STL{} we choose to interpret $\True$ as $1$ and $\False$
as $-1$.
In all these four cases we show that the resulting logic satisfies the
soundness property stated above.
Adding $-\infty$ and~$+\infty$ has repercussions when proving the
geometric properties of the logics, as we show later in
Sect.~\ref{sec:differentiability}. 
If it were not for considerations of using the generic 
syntax for all DLs, 
$\True$ and $\False$ could be removed from the \STL{} syntax altogether, 
without damaging the main results.

\subsubsection*{Lessons learnt}
Soundness for DLs with open intervals was the first real challenge that this formalisation faced.
Having no plausible solution in the field, being able to use \coq{} to 
experiment with different soundness statements and see their  effect on proofs 
was extremely rewarding. Overall, we proved three different versions of 
\STL{} soundness (one for ``vacuous proofs'', which we do not present here);
and we intend to use this formalisation to experiment further with \STL. 
In particular, finding an alternative approach to negation, e.g., using 
``approximate $0$'',
is now within our reach. 
The currently presented approach is the first proof of soundness for any fragment
of \STL{}, it already covers formalisation of problems such as the $\epsilon$-$\delta$-robustness;
 and we attribute this intermediate success to the assistance of the \coq{} formalisation.

\section{Differentiability: shadow-lifting}
\label{sec:differentiability}

It was Varnai et al.\ who provided for \STL{} the pencil-and-paper
proof of shadow-lifting~\cite[Sect.~V]{varnai} (along with the
definition of the \STL{} conjunction). This section formalises this
result and actually completes it since the original proof only covers
one of the two non-trivial cases. The main technical aspect of the
proof is high-school level mathematics: an application of
L'H\^opital's rule, which was not yet available in \analysis.

\DLtwo{} and the \product{} DL also trivially enjoy shadow-lifting.
In the following, we will therefore start by formalising the latter
DLs, then formalising L'H\^opital's rule, and finally provide an
overview of the missing part of Varnai et al.'s proof of
shadow-lifting for \STL{}.
Note that the logics \Godel{}, \Luka{}, \Yager{} fail shadow-lifting
as they are not differentiable everywhere, due to their use of $\min$
or $\max$ to define conjunction.

\subsection{formalisation of shadow-lifting}
\label{sec:shadow-lifting}

As seen in Sect.~\ref{sec:background},
shadow-lifting is defined in terms of partial derivatives, for which
there was however no theory yet in \analysis{}. They can be
easily defined on the model of derivatives \cite[file \coqin{derive.v}]{analysis}.
First, we define \newterm{error vectors} as row vectors (type \coqin{'rV[R]_k} where \coqin{R} is some ring)
that are $0$ everywhere except at one coordinate \coqin{i}:
\begin{minted}[fontsize=\footnotesize]{ssr}
Definition err_vec {R : ringType} (i : 'I_n.+1) : 'rV[R]_n.+1 := 
  \row_(j < n.+1) (i == j)%:R.
\end{minted}
The notation \mintinline{coq}{
note that here the boolean equality (notation: \coqin{==}) is implicitly coerced to a natural number.
Then, given a function \coqin{f} that takes as input a row vector, we
define a function \coqin{partial} that given a row vector \coqin{a}
and an index~\coqin{i} returns the limit
$\lim_{\substack{h\to 0\\h\neq 0}}\frac{\texttt{f}(\texttt{a} +
  h\texttt{err_vec i}) - \texttt{f}(\texttt{a})}{h}.$
Put formally:
\begin{minted}[fontsize=\footnotesize]{ssr}
Definition partial {R} {n} (f : 'rV[R]_n.+1 -> R) (a : 'rV[R]_n.+1) i :=
  lim (h^-1 * (f (a + h *: err_vec i) - f a) @[h --> 0^']).
\end{minted}
In this syntax, \coqin{0^'} represents the deleted neighborhood of \coqin{0},
and \coqin{lim g @ F} represents the limit of the function \coqin{g} at the 
filter \coqin{F} \cite{affeldt2018formalization}.
The notation \coqin{*:} represents scaling but is equivalent to the multiplication of real numbers here.
Hereafter, we use the \coq{} notation \coqin{d f '/d i} for \coqin{partial f i}.

Using partial derivatives, the definition of shadow-lifting (Definition~\ref{def:shadow}) translates directly into \coq:
\begin{minted}[fontsize=\footnotesize]{ssr}
Definition shadow_lifting {R : realType} n (f : 'rV_n.+1 -> R) :=
  forall p, p > 0 -> forall i, ('d f '/d i) (const_mx p) > 0.
\end{minted}
The \coqin{const_mx} function comes from \mathcomp's matrix theory and represents a matrix where all coefficients are the given constant; we use it to implement the restriction ``$p_j = p$'' seen in Definition~\ref{def:shadow}.

\subsection{Shadow-lifting for \DLtwo{} and \product{}}
\label{sec:sldltwoproduct}

The proof of shadow-lifting for \DLtwo{} and \product{} DLs provides
an easy illustration of the use of the definition of the previous section (Sect.~\ref{sec:shadow-lifting}).

For \DLtwo{}, the first thing to observe is that the semantics of a
vector of real numbers can simply be written as an iterated sum
using the notation \coqin{``_} to address elements of row-vectors:
\begin{minted}[fontsize=\footnotesize]{ssr}
Definition dl2_and {R : fieldType} {n} (v : 'rV[R]_n) := \sum_(i < n) v ``_ i.
\end{minted}
Shadow-lifting for \DLtwo{} really just amounts to checking that the
partial derivatives of the function
$\vec{v}\mapsto \sum_{j < |\vec{v}|}\vec{v}_j$ are 1, i.e.,
considering vectors of size \coqin{M.+1}:
\begin{minted}[fontsize=\footnotesize]{ssr}
Lemma shadowlifting_dl2_andE (p : R) : p > 0 ->
  forall i, ('d (@dl2_and R M.+1) '/d i) (const_mx p) = 1.
\end{minted}
Since the partial derivatives are all positive, \DLtwo{} satisfies the
\coqin{shadow_lifting} predicate, see \cite[file \coqin{dl2.v}]{ldl}.

Similarly, we observe for the \product{} DL that the semantics of a vector
is the function $\vec{v}\mapsto \prod_{j < |\vec{v}|}\vec{v}_j$ whose 
partial derivatives are $p^M$, which is positive:
\begin{minted}[fontsize=\footnotesize]{ssr}
Lemma shadowlifting_product_andE p : p > 0 ->
  forall i, ('d (@product_and R M.+1) '/d i) (const_mx p) = p ^+ M.
\end{minted}

\subsection{formalisation of L'H\^opital's rule using \analysis{}}
\label{sec:lhopital}

As indicated in the introduction of this section, the key technical
lemma to prove shadow-lifting for \STL{} is L'H\^opital's rule, that
we show how to formalise in \analysis{}. As a reminder, here follows
one of L'H\^opital's rules:
\begin{theorem}[L'Hôpital's rule]
Let $f, g: \Real \to \Real$ be functions differentiable on an open interval $U$ 
except possibly at one point $a$. Suppose that $\forall x \in U$, $x\neq a$, we have
$g'(x) \neq 0$. If it holds that $f(a)=g(a)=0$, then
if $\lim\limits_{x \rightarrow a^+} \frac{f'(x)}{g'(x)} = l$ for some real number $l$,
then $\lim\limits_{x \rightarrow a^+} \frac{f(x)}{g(x)} = l$.
\end{theorem}
It can be formally stated with \analysis{} using: (a)~the relation
\coqin{is_derive} (between a function and its derivative at some
point: the \coqin{1} appearing in the \coqin{is_derive} expression is
the direction of the derivative, which is $1$ for real number-valued
functions) and (b)~the \newterm{right filter} \coqin{a^'+} (i.e., the
filter of neighborhoods of \coqin{a} intersected with
$]\texttt{a},+\infty]$). We slightly generalize the above statement by
considering a neighborhood \coqin{U} of \coqin{a} instead of an open (line~\ref{line:nbhs})
and by having the derivative of \coqin{g} non-zero ``near'' \coqin{a} (line~\ref{line:nearneq0})
\cite[Sect.~3.2]{affeldt2018formalization}:
\begin{minted}[fontsize=\footnotesize,numbers=left,xleftmargin=1.5em,escapeinside=88]{ssr}
Context {R : realType}.
Variables (f df g dg : R -> R) (a : R) (U : set R) (Ua : nbhs a U). 8\label{line:nbhs}8
Hypotheses (fdf : forall x, x \in U -> is_derive x 1 f (df x))
           (gdg : forall x, x \in U -> is_derive x 1 g (dg x)).
Hypotheses (fa0 : f a = 0) (ga0 : g a = 0)
           (cdg : \forall x \near a^', dg x != 0). 8\label{line:nearneq0}8
Lemma lhopital_right (l : R) :
  df x / dg x @[x --> a^'+] --> l -> f x / g x @[x --> a^'+] --> l.
\end{minted}
The proof is textbook, relying in particular on Cauchy's Mean
  Value Theorem (MVT), the proof of which can be derived from the
already available MVT, see \cite[file \coqin{stl.v}]{ldl}.
Note that we also need the variant for the left filters.

\subsection{Shadow-lifting for \STL{}}
\label{sec:slstl}

Compared with \DLtwo{} and \product{}, the conjunction of \STL{}
($and_S$ in Table~\ref{tab:semantics-math}) is much more involved: it
consists of two non-trivial cases (marked as $a_{\min} < 0$ and $a_{\min} > 0$
in Table~\ref{tab:semantics-math}) whose computation requires
summations of exponentials of deviations.
Varnai et al.  provide a proof sketch for the case
$a_{\min} > 0$~\cite[Sect.~V]{varnai} which we have successfully
formalised, using in particular l'H\^opital's rule from the previous
section. Below we explain the formalisation of the other case
$a_{\min} < 0$ that Varnai et al.\ did not treat.

The case $a_{\min} < 0$ actually refers to the semantics provided
by the function \coqin{stl_and_lt0} already presented in
Fig.~\ref{fig:semantics-lt0-gt0}.
The positive limit we are looking for is actually $\frac{1}{M+1}$
(where $M+1$ is the size of vectors), i.e., our goal is to prove formally
(the notation \coqin{\o} is for function composition):
\begin{minted}[fontsize=\footnotesize]{ssr}
Lemma shadowlifting_stl_and_lt0 (p : R) : p > 0 -> forall i,
  ('d (stl_and_lt0 \o seq_of_rV) '/d i) (const_mx p) = M.+1%:R^-1.
\end{minted}
This boils down to proving the existence of the limit ``from below''
and ``from above''.
The ``from below'' case consists in the following convergence lemma:
\begin{minted}[fontsize=\footnotesize]{ssr}
Lemma shadowlifting_stl_and_lt0_cvg_at_left (p : R) i : p > 0 ->
  h^-1 * (stl_and_lt0 (seq_of_rV (const_mx p + h *: err_vec i)) -
          stl_and_lt0 (seq_of_rV (const_mx p))) @[h --> 0^'-] --> M.+1%:R^-1.
\end{minted}
For the sake of clarity, let us switch to standard mathematical
notations and assume without loss of generality that \coqin{i} is actually \coqin{M}.
By mere algebraic transformations (using \mathcomp's algebra theory),
the goal can be turned into a sum of two limits:
$$
\begin{array}{ll}
\lim\limits_{h \rightarrow 0^-} \dfrac{\tSTL{\naryConj(\pval, \ldots, \pval, 
		\pval + h)} - \tSTL{\naryConj(\pval, \ldots, \pval)}}{h} \\ 
= \lim\limits_{h \rightarrow 0^-} \frac{1}{h} \left(
	\dfrac{(\pval + h) M e^{\frac{-h}{\pval+h}} e^{\nu 
	\frac{-h}{\pval+h}} + \pval + h }{Me^{\nu \frac{-h}{\pval+h}} + 1}
	- p
	\right) & \text{by definition (see Table~\ref{tab:semantics-math})}\\
= \underbrace{\lim\limits_{h \rightarrow 0^-} \frac{h}{h(M + e^{\nu \frac{h}{\pval+h}})}}_{(a)}
 +
\underbrace{\lim\limits_{h \rightarrow 0^-} \frac{M (\pval + 
h)e^{\frac{-h}{\pval+h}} - \pval M }{h(M + e^{\nu \frac{h}{\pval+h}})}}_{(b)} & 
\text{by simplification} \\
\end{array}
$$
We can show directly that $(a)=\frac{1}{M+1}$
but the computation of $(b)$ requires L'H\^opital's rule:
$$
\begin{array}{rcl}
(b) & = & \lim\limits_{h \rightarrow 0^-} \frac{\frac{h M 
	e^{\frac{-h}{\pval+h}}}{\pval + h}}
	{e^{\nu \frac{h}{\pval+h}} +
	h e^{\nu\frac{-h}{\pval+h}}(\frac{\nu}{\pval + h} - \frac{h \nu}{(\pval 
	+h)^2}) + M} \\
& = & \lim\limits_{h \rightarrow 0^-} h 
	\lim\limits_{h \rightarrow 0^-} \frac{M e^{\frac{-h}{\pval+h}}}{\pval + h}
	\lim\limits_{h \rightarrow 0^-} \frac{1}{e^{\nu \frac{h}{\pval+h}} +
		h e^{\nu\frac{-h}{\pval+h}}(\frac{\nu}{\pval + h} - \frac{h 
		\nu}{(\pval 
			+h)^2}) + M} \\
& = & 0 \cdot \frac{M}{\pval} \cdot \frac{1}{1 + M} = 0 \\
\end{array}
$$
 
Barring the necessity of finding the most convenient breakdown of the
limit in the two penultimate steps, this proof is arguably
mathematically straightforward.  The corresponding mechanised proof is
however significantly less trivial than in the cases of \product{} and
\DLtwo{} (Sect.~\ref{sec:sldltwoproduct}): length-wise the first
tentative formal proof we wrote was an order of magnitude larger.

Proving the ``from above'' above consists of a simpler but similar
argument:
\begin{multline*}
\lim\limits_{h \rightarrow 0^+} \dfrac{\tSTL{\naryConj(\pval, \ldots, \pval, 
		\pval + h)} - \tSTL{\naryConj(\pval, \ldots, \pval)}}{h}
= \lim\limits_{h \rightarrow 0^+}  \frac{1}{h} \left (\frac{\pval M + 		
e^{\frac{h}{\pval}} e^{\frac{\nu h}{\pval}}} {M + e^{\frac{h}{\pval}}} - p\right) \\
= \lim\limits_{h \rightarrow 0^+} \frac{1}{M + e^{\frac{\nu h}{\pval}}}
\lim\limits_{h \rightarrow 0^+} e^{\frac{\nu h}{\pval}}
\lim\limits_{h \rightarrow 0^+} \frac{ e^{\frac{ h}{\pval}} 
-1}{\frac{h}{\pval}}
= \frac{1}{M + 1} \cdot 1 \cdot 1 = \frac{1}{M + 1}\\
\end{multline*}

Combined with the formalisation of the case $a_{\min} > 0$
sketched by Varnai et al.\ in~\cite[Sect.~V]{varnai}, this completes
the formal proof of shadow-lifting for \STL{}.

\section{Conclusions, related and future work}
\label{sec:conclusion}

We have presented a complete \coq{} formalisation of a range of
existing DLs; making the following two main contributions:

\begin{enumerate}
\item We contribute to the DL community by {\em revisiting
    semantics of \STL{} and \DLtwo\/} in a way more amenable to formal
  verification. We find and fix errors in the literature.
\item We propose a {\em general formalisation strategy} based on
  dependent types and formal mathematics. The proposed
  formalisation is built to be easily extendable for future studies of
  different DLs.
\end{enumerate}

\begin{table}
\centering
\begin{tabular}{lll}
File & Contents & L.o.c. \\
\hline
\multicolumn{3}{|l|}{\it Additions to \mathcomp{} libraries} \\
\hline
{\tt mathcomp\us{}extra.v} & Lemmas iterated $\min$/$\max$, etc. & 504 \\
{\tt analysis\us{}extra.v} & L'H\^opital's rule, Cauchy's MVT (\S~\ref{sec:lhopital}), etc. & 820 \\
\hline
\multicolumn{3}{|l|}{\it Generic logic and generic definitions of properties } \\
\hline
{\tt ldl.v} & LDL syntax and semantics(\S~\ref{sec:syntax}), shadow-lifting (Sect.~\ref{sec:shadow-lifting}) & 417 \\
\hline
\multicolumn{3}{|l|}{{\it Soundness, logical and geometric properties of concrete logics}} \\
\hline
{\tt dl2.v} & \DLtwo{}: logical (\S~\ref{sec:logical}), geometric (\S~\ref{sec:sldltwoproduct}) & 259  \\
{\tt fuzzy.v} & \Godel, \Luka, \Yager, \product:  & 731  \\
& logical (\S~\ref{sec:logical}), geometric (\S~\ref{sec:sldltwoproduct}) & \\
{\tt stl.v} & \STL{}; logical (\S~\ref{sec:logical}), geometric (\S~\ref{sec:slstl}) & 977 \\
\hline
\multicolumn{3}{|l|}{{\it Alternative formalisations of logical properties/ 
soundness using extended reals }} \\
\hline
{\tt dl2\us{}ereal.v} & \DLtwo{}: logical (\S~\ref{sec:sound}) & 211\\
{\tt stl\us{}ereal.v} & \STL{}: logical (\S~\ref{sec:sound})& 362\\
& \multicolumn{1}{r}{Total} & 4281 \\
\end{tabular}
\caption{Overview of the formalisation~\cite{github}}
\label{tab:overview}
\end{table}

Table~\ref{tab:overview} summarises the \coq{} implementation. 
Both L'Hôpital rule and geometric properties, especially in complex cases such 
as \STL{}, form a substantial part of the development. 
The proofs for fuzzy DLs (file \coqin{fuzzy.v}) are grouped together as
thanks to their similarity, they share some of the proofs. The files
\coqin{mathcomp_extra.v} and \coqin{analysis_extra.v} contain utility
lemmas for the respective libraries. The file \coqin{mathcomp_extra.v}
has a selection of lemmas on big operations (e.g., iterated sums, $n$-ary
maximum), including lemmas for said operations when restricted to the
domain $[0,1]$ used by fuzzy logics. The file \coqin{analysis_extra.v}
on the other hand contains proofs of L'Hôpital, Cauchy's MVT, and
multiple lemmas on properties of \coqin{mine} and \coqin{maxe} ($\min$
and $\max$ for extended reals).

During our work, we completed missing parts of the shadow-lifting
proof for \STL{}, for example, the original \STL{} proof failed to
acknowledge the need for L'Hôpital rule.
We believe that understanding, let alone verifying theories that
pertain to AI, without any mechanised support is difficult.
Our previous attempt~\cite{ldl} to do this with pen and paper proofs
was drowned in low-level case analysis and resulted in some errors,
see Example~\ref{ex:errors}.  This complexity was our initial
motivation to undertake the formalisation.

Regarding the concrete formalisation strategy, it was revealing that
most of our formalisation was coherent with the standard \mathcomp{}
libraries (and standard mathematical results), and the library
extensions we needed were natural (e.g., L'Hôpital rule).  This
work hence demonstrates that \coq{} and \mathcomp{} are effective
working tools to formalise state-of-the-art AI results: \DLtwo{} and \STL{}
were published in recent conferences\cite{varnai,fischer2019dl2} and
this paper formalises the most significant results from both.

In the end, we have a uniform formalisation where all DLs are “tamed” which 
provides 
solid ground for formalisation of methods deployed in verification of neural 
networks.

\paragraph*{Related Work} 
As this paper illustrates, neural network verification is a new field,
and its nascent methods need validation and further refinement.

In terms of programming language support for neural network verification, 
a tool CAISAR~\cite{girardsatabin2022caisar}, implemented as an OCaml DSL, 
puts emphasis on the smooth integration of a general specification language 
with many existing neural network solvers. 
However, CAISAR does not support property-guided training.  
The aspiration of languages like Vehicle and CAISAR is to accommodate compilation
of specifications into both neural network solver and machine learning backends.
For the former, there is an on-going work on certifying the neural network 
solver backends~\cite{DesmartinIPSKK23,DaggittAKKA23}. 

On the side of machine-learning backends,
DLs have been previously formalised in Agda as part of the Vehicle formalisation~\cite{ADK24}, 
but did not extend to shadow-lifting---the part that requires extensive mathematical libraries. 
Property-guided training certified via theorem proving was also proposed in~\cite{ChevallierWF22}. 

Relevant work on formalisation of neural networks in ITPs includes:
verification of neural networks in Isabelle/HOL
\cite{brucker2023verifying} and Imandra~\cite{DesmartinPKD22},
formalisation of piecewise affine activation functions in
\coq~\cite{aleksandrov2023formalizing}, providing formal guarantees of
the degree to which the trained neural network will generalise to new
data in \coq~\cite{bagnall2019certifying}, convergence of a
single-layered perceptron in \coq~\cite{murphy2017verified}; and
verification of neural archetypes in \coq~\cite{de2022use}.  The
formalisation presented here does not directly formalise neural
networks.

\paragraph*{Future Work}
We plan to consider other definitions of soundness, and other DLs,
including \STL{} with revised negation.  We hypothesise that
Definition~\ref{def:shadow} allows for generalisation (removing the
condition ``$p_j = p$'') and this is left for future work.  
We saw in Sect.~\ref{sec:sound} that the choice of the interpretation
domains has an impact on both soundness and shadow-lifting and this
choice might deserve further investigation.
The trade-off between idempotence, associativity and shadow-lifting
that was conjectured in~\cite{varnai} is reminiscent of substructural
logics and suggests investigating the connection.
Establishing connection of this work with the logics of Lawvere
quantale by Bacci et al.\ \cite{abs-2302-01224} 
might also provide new tools to study DLs.

Separately from the questions of scientific curiosity and mathematical elegance,
there is a question of lacking programming language support for machine learning. 
As tools like Vehicle~\cite{daggitt2024vehicle} and CAISAR~\cite{girardsatabin2022caisar}
 are being proposed to provide a more principled 
approach to verification of machine learning, in the long term, compilers of 
these new emerging languages will require certification. And this, in turn, will demand formalisation
of results such as the ones we presented here. The formalisation of DLs would hence directly contribute to 
certified compilation of specification languages to machine learning libraries.   



\bibliography{bibliography}

\end{document}